\documentclass[12pt,a4paper]{article}
\usepackage[reqno]{amstex}
\usepackage{amscd}
\newcommand{\bilin}[2]{\left\langle{#1},{#2}\right\rangle}
\newcommand{\binox}[2]
   {\begin{pmatrix}{#1}\\{#2}\end{pmatrix}}
\newcommand{\romaC}{{\mathrm C}}
\newcommand{\dbeta}{\dot{\beta}}
\newcommand{\dfunc}[1]{\frac{\delta}{\delta{#1}}}
\newcommand{\dilat}[1]
   {\left({\mathcal D}{#1},\frac{\delta{#1}}{\delta{#1}}\right)}
\newcommand{\dmuGa}[1]{{\rm d}\mu_{\Gamma_L}(#1)}
\newcommand{\dpart}[1]{\frac{\partial}{\partial{#1}}}
\newcommand{\dmom}[2]{\frac{{\mathrm d}^{{#1}}{#2}}{(2\pi)^{{#1}}}}
\newcommand{\dscale}[1]{\frac{{\mathrm d}{#1}}{{#1}}}
\newcommand{\dtota}[1]{\frac{{\mathrm d}}{{\mathrm d}{#1}}}
\newcommand{\intex}[1]
   {\int{\mathrm d}^Dx_1\cdots{\mathrm d}^Dx_{{#1}}}
\newcommand{\intep}[1]
   {\int\frac{{\mathrm d}^Dp_1}{(2\pi)^D}\cdots
   \frac{{\mathrm d}^Dp_{{#1}}}{(2\pi)^D}}
\newcommand{\KK}{{\mathcal K}}
\newcommand{\kkern}[2]
   {{\mathcal K}^{({#1})}_{#2}(x_1,\ldots,x_{{#2}})}
\newcommand{\N}{{\mathbb N}}
\newcommand{\obs}[2]{{\mathcal O}_{{#1}}({#2})}
\newcommand{\pertv}[2]{{\mathcal V}^{({#1})}({#2})}
\newcommand{\pertb}[1]{{\dot{\beta}}^{({#1})}}
\newcommand{\pertk}[2]{{\mathcal K}^{({#1})}({#2})}
\newcommand{\pertsp}[2]{\widetilde{\sigma}^{({#1})}_{{#2}}}
\newcommand{\pertsr}[2]{\sigma^{({#1})}_{{#2}}}
\newcommand{\pertz}[1]{\zeta^{({#1})}}
\newcommand{\phixs}[1]{\phi(x_1)\cdots\phi(x_{{#1}})}
\newcommand{\pkker}[2]
   {\widetilde{{\mathcal K}}^{({#1})}_{{#2}}
   (p_1,\ldots,p_{{#2}})}
\newcommand{\pkern}[2]
   {\widetilde{{\mathcal V}}^{({#1})}_{{#2}}
   (p_1,\ldots,p_{{#2}})}
\newcommand{\poten}[2]{{\mathcal V}({#1}\vert{#2})}
\newcommand{\pro}[1]{{{\mathcal P}}_{{#1}}}
\newcommand{\R}{{\mathbb R}}
\newcommand{\RG}{{\mathcal R}}
\newcommand{\rkern}[2]
   {{\mathcal V}^{({#1})}_{#2}(x_1,\ldots,x_{{#2}})}
\newcommand{\rkker}[2]
   {{\mathcal K}^{({#1})}_{#2}(x_1,\ldots,x_{{#2}})}
\newcommand{\SG}{{\mathcal S}}
\newcommand{\VV}{{\mathcal V}}
\newcommand{\wt}[1]{\widetilde{{#1}}}
\newcommand{\Z}{{\mathbb Z}}

\begin{document}
\begin{titlepage}
\begin{center}
{\Large The renormalized $\phi^4_4$-trajectory by \\[2mm]
perturbation theory in a running coupling II:\\[2mm]
the continuous renormalization group} \\[10mm]
\end{center}
\begin{center}
{\Large C. Wieczerkowski} \\[10mm]
\end{center}
\begin{center}
Institut f\"ur Theoretische Physik I,
Universit\"at M\"unster, \\
Wilhelm-Klemm-Stra\ss e 9, D-48149 M\"unster, \\
wieczer@@ uni-muenster.de \\[2mm]
\end{center}
\vspace{-10cm}
\hfill MS-TP1-96-14
\vspace{11cm}
\begin{abstract}
The renormalized trajectory of massless $\phi^4$--theory on
four dimensional Euclidean space--time is investigated as a
renormalization group invariant curve in the center manifold
of the trivial fixed point, tangent to the $\phi^4$--interaction.
We use an exact functional differential equation for its
dependence on the running $\phi^4$--coupling. It is solved by
means of perturbation theory. The expansion is proved to be
finite to all orders. The proof includes a large momentum bound
on amputated connected momentum space Green's functions.
\end{abstract}
\end{titlepage}

\section{Introduction}

In Wilson's renormalization group \cite{W71,WK74}, renormalized
theories come as renormalized trajectories of effective actions.
The renormalization group leaves invariant a renormalized
trajectory up to a flow of renormalization parameters. This
invariance prescribes a renormalized theory without a detour to
limit procedures.

Renormalization group invariance was used in \cite{Wi96} as a
first principle to deduce a renormalized running coupling
expansion for the renormalized trajectory of massless
$\phi^4$--theory on four dimensional Euclidean space--time.
The construction yielded a pair consisting of an effective
potential $V(\phi|g)$ and a step $\beta$--function $\beta_L(g)$,
both in perturbation theory in $g$, which obey the discrete
flow equation
\begin{equation}
\RG_L V(\phi|g)=V(\phi|\beta_L(g)).
\label{1.1}
\end{equation}
Here $\RG_L$ stands for a momentum space renormalization group
transformation, based on the decomposition of
$(-\bigtriangleup)^{-1}$ by means of an exponential regulator,
which rescales by a factor of $L>1$. Eq. (\ref{1.1}) was shown
to possess a unique solution consisting to first order in $g$
of a $\phi^4$--vertex together with a wave function term,
\begin{equation}
V(\phi|g)=g\left\{\frac{1}{4!}\int{\rm d}^Dx\;\phi(x)^4+
\frac{\zeta^{(1)}}{2}\int{\rm d}^Dx\;\phi(x)
(-\bigtriangleup)\phi(x)\right\}+O(g^2),
\label{1.2}
\end{equation}
where $D=4$, and consisting to all higher orders of polynomial
vertices with all properties of a renormalized theory. The higher
orders of (\ref{1.2}) where determined inductively. The step
$\beta$--function came out
as
\begin{equation}
\beta_L(g)=g-\frac{3\log (L)}{(4\pi )^2}g^2+O(g^3),
\label{1.3}
\end{equation}
showing asymptotic freedom for the forward flow on the
renormalized $\phi^4$--trajectory. The construction relied on
the analysis of $\RG_L$ with a fixed scale $L>1$.

The scale parameter $L$ is the ratio of an ultraviolet and an
infrared cutoff. Remarkably, the effective potential
$V(\phi|g)$ comes out independent of $L$. The only
$L$--dependence remains in the step $\beta$--function (\ref{1.3}).
The differential $\beta$--function
\begin{equation}
\dbeta (g)=L\frac{{\rm d}}{{\rm d}L}\beta_L (g)=
\frac{-3}{(4\pi)^2}g^2+O(g^3)
\label{1.4}
\end{equation}
on the other hand comes out independent of $L$. This
suggests a differential approach for the pair $V(\phi|g)$
and $\dbeta (g)$ which makes no reference to the scale $L$
at all. That is the present program. It yields an
alternative construction for the renormalized
$\phi^4$--trajectory.

The differential form of the discrete flow equation
(\ref{1.1}) is a functional differential equation. For the
normal ordered potential $V(\phi\vert g)=:\poten{\phi}{g}:$
it reads
\begin{equation}
\left\{\dbeta(g)\dpart{g}-\dilat{\phi}\right\}
\poten{\phi}{g}=
-\bilin{\poten{\phi}{g}}{\poten{\phi}{g}}.
\label{1.5}
\end{equation}
The operator $\dilat{\phi}$ generates scale transformations.
The right hand side of (\ref{1.5}) is a bilinear
renormalization group bracket $\bilin{A(\phi)}{B(\phi)}$.
It consists of contractions between $A(\phi)$ and $B(\phi)$,
and is independent of $g$.

Restricted to the renormalized trajectory, the renormalization
group becomes a one--dimensional dynamical system given by
a flow of $g$ according to the ordinary differential equation
\begin{equation}
L\frac{{\rm d}}{{\rm d}L}g(L)=\dbeta (g(L)).
\label{1.6}
\end{equation}
The step $\beta$--function (\ref{1.3}) is retained as a
solution of (\ref{1.6}).

The differential approach parallels completely its discrete
brother \cite{Wi96}. It yields an alternative construction of
the same renormalization group invariant. The scheme will be
to  solve (\ref{1.5}) inductively by perturbation theory in
$g$. The $\phi^4$--trajectory will be selected as the unique
solution which is given by (\ref{1.2}) to first order in $g$,
and higher orders of appropriate type. We remark that the first
order wave function renormalization in (\ref{1.2}) is peculiar
to the four dimensional case.

Both the aim and the setup of this paper are identical to
those in \cite{Wi96}. However, the difference equations of
\cite{Wi96} are here replaced by differential equations.
The formulation to be preferred is a matter of taste. The
differential version has the advantage that it involves a
quadratic non--linearity only. As a consequence, the
inductive solution involves only pairwise convolution of
lower order vertices.

\section{Renormalization group}

The setup will be as in \cite{Wi96}, with the difference that
the scale $L$ is here variable.

Consider the following momentum space renormalization group
transformation $\RG_L$, depending on a scale parameter $L>1$.
Let $\RG_L$ be composed of a Gaussian fluctuation integral,
with covariance $\Gamma_L$ and mean $\psi$, and a dilatation
$\SG_L$ of $\psi$. Let the fluctuation covariance be defined by
\begin{equation}
\wt{\Gamma}_L(p)=\frac 1{p^2}
\bigl\{\wt{\chi}(p)-\wt{\chi}(Lp)\bigr\},
\label{2.1}
\end{equation}
where $\wt{\chi}(p)=\exp (-p^2)$. Let $\dmuGa{\zeta}$ be the
associated Gaussian measure on field space with mean zero.
Let $\SG_L$ be defined by $\SG_L\psi(x)=L^{1-D/2}\psi(x/L)$.
We then define $\RG_L$ as the renormalization group transformation
\begin{equation}
\RG_LV(\psi)=-\log\int\dmuGa{\zeta}\exp
\left\{-V(\SG\psi+\zeta)\right\}.
\label{2.2}
\end{equation}
We will restrict our attention to $\Z_2$--symmetric potentials,
with $V(-\phi)=V(\phi)$. The field independent constant will be
understood to be removed in (\ref{2.2}).

The composition of two renormalization group transformations with
scale $L$ is equal to one renormalization group transformation
with scale $L^2$. Moreover, the transformation (\ref{2.2}) satisfies
\begin{gather}
\RG_{L_1} \RG_{L_2}=\RG_{L_1 L_2}, \quad L_1, L_2>1,
\label{2.3} \\
\lim_{L\rightarrow 1^+} \RG_L = {\mathrm id}.
\label{2.4}
\end{gather}
It therefore defines a representation of the semi--group of
dilatations with scale factors $L>1$ on the space of effective
potentials.

Due to the semi-group property the iteration of renormalization
group transformations with fixed scale is identical with an
increase of the scale in a single transformation. This
interpolation is the motive of the present investigation, in
conjunction with the previous construction in \cite{Wi96}.
The continuous point of view has the advantage to allow for
infinitesimal renormalization group transformations, which can
be expected to be close to the identity.

The renormalization group transformation (\ref{2.2}) associates
with a bare potential $V(\phi)$ an orbit $V(\phi|L)=\RG_L V(\phi)$.
This orbit satisfies the functional differential equation
\begin{eqnarray}
&&\left\{ L\dpart{L}-
\dilat{\phi}-
\frac{1}{2}\left(\frac{\delta}{\delta\phi},
\romaC\frac{\delta}{\delta\phi}\right)
\right\} V(\phi|L)=
\nonumber\\&&\qquad
\frac{-1}{2} \left(\frac{\delta}{\delta \phi}V(\phi|L),\romaC
\frac{\delta}{\delta \phi}V(\phi|L)\right).
\label{2.5}
\end{eqnarray}
Here ${\mathcal D}\phi(x)=(1-D/2-x\partial_x)\phi(x)$
generates dilatations of the field $\phi$. The operator
$\left({\mathcal D}\phi,\delta_{\phi}\right)$ generates
scale transformations of the potential,
\begin{equation}
\dilat{\phi}V(\phi)=\dtota{L}V(\SG_L\phi)
\biggr\vert_{L=1}.
\label{2.6}
\end{equation}
The operator $\romaC$ is given by
\begin{equation}
\romaC=\SG_{L^{-1}}
\left(L\dpart{L}\Gamma_L\right)
\SG_{L^{-1}}^T,
\label{2.7}
\end{equation}
and is $L$--independent. With an exponential regulator we have
that $\romaC=2\chi$. Eq. (\ref{2.5}) follows from a functional
heat equation for convolutions by a parameter dependent
Gaussian measure, see \cite{GJ87}.

The functional Laplacian on the left hand side of (\ref{2.5})
can be removed by normal ordering at the expense of a more
complicated bilinear term. We define a normal ordered potential
$\poten{\phi}{L}$ by
\begin{equation}
V(\phi|L)=\exp\left\{\frac{-1}{2}
\left(\frac{\delta}{\delta\phi},v
\frac{\delta}{\delta\phi}\right)\right\}
\poten{\phi}{L}.
\label{2.8}
\end{equation}
Strictly speaking $\poten{\phi}{L}$ is the pre--image of
$V(\phi|L)$ under the normal ordering operator and is
therefore not decorated by normal ordering colons. To remove
the functional Laplacian, the normal ordering covariance has
to be chosen such that
\begin{equation}
\dtota{L}\SG_{L^{-1}}v(\SG_{L^{-1}})^T
\biggr\vert_{L=1}=\romaC.
\label{2.9}
\end{equation}
Specifically, we choose
\begin{equation}
\wt{v}(p)=\frac{\exp(-p^2)}{p^2},
\label{2.10}
\end{equation}
a massless covariance with unit ultraviolet cutoff.
The normal ordered potential then satisfies the functional
differential equation
\begin{equation}
\left\{L\dpart{L}-\dilat{\phi}\right\}
\poten{\phi}{L}=
-\bilin{\poten{\phi}{L}}{\poten{\phi}{L}}.
\label{2.11}
\end{equation}
The non--linearity comes in form of a bilinear renormalization
group bracket
\begin{eqnarray}
&&\bilin{A(\phi)}{B(\phi)}=
\nonumber\\&&\qquad
\frac{1}{2}\left(\dfunc{\phi_1},\romaC\dfunc{\phi_2}\right)
\exp\left\{\left(\dfunc{\phi_1},v\dfunc{\phi_2}\right)\right\}
A(\phi_1)\;B(\phi_2)\biggr\vert_{\phi_1=\phi_2=\phi},
\label{2.12}
\end{eqnarray}
where $\phi_1$ and $\phi_2$ denote two independent copies of
$\phi$. It consists of contractions between $A(\phi)$ and
$B(\phi)$, each contraction being made of one hard propagator
$\romaC$ and any number of soft propagators $v$.

\section{$\phi^4$--trajectory}

Consider the following renormalization problem. We seek
an effective potential $\poten{\phi}{g}$ and a differential
$\beta$--function $\dbeta(g)$, both depending on a coupling
parameter $g$, but not on $L$, with the following
properties of a renormalized theory.\\[2mm]

\noindent{\bf 1. Power series:} the effective potential
$\poten{\phi}{g}$ and the differential $\beta$--function
$\dbeta(g)$ are both formal power series
\begin{eqnarray}
\poten{\phi}{g}&=&
\sum_{r=1}^\infty\frac{g^r}{r!}\pertv{r}{\phi},
\label{3.1}
\\
\dbeta(g)&=&
\sum_{r=1}^\infty\frac{g^r}{r!}\pertb{r},
\label{3.2}
\end{eqnarray}
in $g$. The question of the summability of (\ref{3.1}) and
(\ref{3.2}) will be left aside.\\[2mm]

\noindent{\bf 2. $\phi^4$--theory:} the effective potential
$\poten{\phi}{g}$ is to first order a $\phi^4$--vertex
together with a wave function term,
\begin{equation}
\pertv{1}{\phi}=
\frac{1}{4!}\int{\rm d}^Dx\;\phi(x)^4+
\frac{\pertz{1}}{2}\int{\rm d}^Dx\;\phi(x)
(-\bigtriangleup)\phi(x),
\label{3.3}
\end{equation}
where $\pertz{1}$ is a first order wave function parameter.
The $r$'th order effective potential is a polynomial
\begin{equation}
\pertv{r}{\phi}=
\sum_{m=1}^{r+1}\frac{1}{(2m)!}\intex{2m}
\;\phixs{2m}\; \rkern{r}{2m}
\label{3.4}
\end{equation}
in the field $\phi$.\\[2mm]

\noindent{\bf 3. Regularity:} the kernels in (\ref{3.4}) are
Euclidean invariant symmetric distributions. They are
given by Fourier integrals
\begin{eqnarray}
&&\rkern{r}{2m}=\intep{2m}
\exp\left(i\sum_{l=1}^{2m}p_lx_l\right)
\nonumber\\&&\qquad
(2\pi)^D\delta\left(\sum_{l=1}^{2m}p_l\right)
\pkern{r}{2m}.
\label{3.5}
\end{eqnarray}
With the $\delta$--function removed, their Fourier transforms
are Euclidean invariant symmetric ${\mathcal C}^\infty$--functions
on momentum space $\R^D\times\cdots\times\R^D$. They satisfy
$L_{\infty,\epsilon}$--bounds
\begin{equation}
\|\frac{\partial^{|\alpha |}}{\partial p^\alpha}
\wt{{\mathcal V}}^{(r)}_{2m}\|_{\infty,\epsilon}
<\infty,
\label{3.6}
\end{equation}
for all $\epsilon >0$. Here $\alpha=(\alpha_{l,\mu})\in
\N^D\times\cdots\times\N^D$ is a multi--index and
$|\alpha |=\sum_{l,\mu}\alpha_{l,\mu}$. The
$L_{\infty,\epsilon}$--norm is defined as
\begin{equation}
\|\wt{{\mathcal V}}^{(r)}_{2m}\|_{\infty,\epsilon}=
\sup_{(p_1,\ldots,p_{2m})\in{\mathcal M}_{2m}}
\left\vert\pkern{r}{2m}\right\vert
\exp\left(-\epsilon\sum_{l=1}^{2m}|p_l|\right),
\label{3.7}
\end{equation}
where ${\mathcal M}_{2m}$ denotes the hyperplane of momenta
with $\sum_{l=1}^{2m}p_l=0$.\\[2mm]

\noindent{\bf 4. Coupling parameter:} the four point kernel
at zero momenta is
\begin{equation}
\wt{{\mathcal V}}^{(r)}_{4}(0,0,0,0)=
\delta_{r,1},
\label{3.8}
\end{equation}
zero in higher orders than one.\\[2mm]

\noindent{\bf 5. Invariance:} (\ref{3.1}) and (\ref{3.2})
satisfy the renormalization group equation
\begin{equation}
\left\{\dbeta(g)\dpart{g}-\dilat{\phi}\right\}
\poten{\phi}{g}=
-\bilin{\poten{\phi}{g}}{\poten{\phi}{g}}.
\label{3.9}
\end{equation}
to all orders in $g$.\\[2mm]

The differential $\beta$--function (\ref{3.2}) defines a
running (scale dependent) coupling $g(L)$ as solution to
the ordinary differential equation
\begin{equation}
L\dtota{L}g(L)=\dbeta(g(L)).
\label{3.10}
\end{equation}
The effective potential $\poten{\phi}{g(L)}$, as a function
of $\phi$ and $L$, then satisfies the flow equation
(\ref{2.11}).

In perturbation theory, there exists a unique solution to
this renormalization problem. We will construct it by
induction on the order of $g$. The renormalization group
equation (\ref{3.9}) is independent of $L$. Its perturbative
solution will also be independent of $L$. In other words,
the solution will depend on cutoffs only through a running
coupling. This property is called scaling. A potential
$\poten{\phi}{g}$, together with a differential
$\beta$--function $\dbeta(g)$, which satisfies (\ref{3.9})
is said to scale.

\section{Scaling equations}

Inserting the power series (\ref{3.1}) and (\ref{3.2}) into
(\ref{3.9}), we obtain a system of first order functional
differential equations for their coefficients. It reads
\begin{equation}
\left\{r\pertb{1}-\dilat{\phi}\right\}
\pertv{r}{\phi}=-\pertk{r}{\phi}.
\label{4.1}
\end{equation}
The right hand side of (\ref{4.1}) depends only on lower orders
$\pertv{s}{\phi}$, with $1\leq s\leq r-1$, and is zero to
first order. It is given by
\begin{equation}
\pertk{r}{\phi}=
\sum_{s=2}^{r}\binox{r}{s}\pertb{s}\pertv{r-s+1}{\phi}+
\sum_{s=1}^{r-1}\binox{r}{s}
\bilin{\pertv{s}{\phi}}{\pertv{r-s}{\phi}}.
\label{4.2}
\end{equation}
If $\pertv{s}{\phi}$ is given by (\ref{3.3}) to first order and
is a polynomial (\ref{3.4}) for all orders $2\leq s\leq r-1$, then
also (\ref{4.2}) is a polynomial
\begin{equation}
\pertk{r}{\phi}=
\sum_{n=1}^{r+1}\frac{1}{(2n)!}\intex{2n}
\;\phixs{2n}\;\kkern{r}{2n}.
\label{4.3}
\end{equation}
The polynomial form (\ref{3.4}) is therefore preserved by the
renormalization group equation (\ref{3.9}). Eq. (\ref{4.1})
will be understood as a system of first order partial differential
equations for the kernels in (\ref{3.4}). It reads
\begin{equation}
\left\{\pertsr{r}{2n}+\sum_{m=1}^{2n}x_m\dpart{x_m}\right\}
\rkern{r}{2n}=\rkker{r}{2n}
\label{4.4}
\end{equation}
with real space scaling dimensions
\begin{equation}
\pertsr{r}{2n}=n(2+D)-r\pertb{1}.
\label{4.5}
\end{equation}
We will look for solutions to (\ref{4.4}) whose Fourier transforms
(\ref{3.5}) obey (\ref{3.6}). We therefore switch from real space
to momentum space. In momentum space (\ref{4.4}) becomes a system
of first order partial differential equations
\begin{equation}
\left\{\pertsp{r}{2n}-\sum_{m=1}^{2n}p_m\dpart{p_m}\right\}
\pkern{r}{2n}=\pkker{r}{2n},
\label{4.6}
\end{equation}
with momentum space scaling dimensions
\begin{equation}
\pertsp{r}{2n}=D+n(D-2)-r\pertb{1}.
\label{4.7}
\end{equation}
The contribution $D$ in (\ref{4.7}) comes from translation
invariance. Anticipating $\pertb{1}=4-D$, the scaling
dimension of the $\phi^4$--vertex, the scaling dimensions
(\ref{4.7}) become order independent in four dimensions.
They are given by
\begin{equation}
\pertsp{r}{2n}=4-2n.
\label{4.8}
\end{equation}
The renormalized $\phi^4$--trajectory is a particular solution
to the scaling equation (\ref{4.6}). It is distinguished by the
properties listed above.

\section{Renormalization group PDEs}

We will solve the scaling equations (\ref{4.1}) by induction on
the order of perturbation theory. The induction step consists of
solving a system of first order renormalization group PDEs
(\ref{4.6}) of the general form
\begin{equation}
\left\{p\dpart{p}-\sigma\right\}F(p)=G(p),
\label{5.1}
\end{equation}
where $\sigma\in\Z$, and where $G(p)$ is a given function of
$p\in\R^N$. The integration of (\ref{5.1}) calls for initial
data. It is usually supplied in form of a bare action together
with renormalization conditions at a distinguished scale.
We will instead look for smooth solutions to (\ref{5.1}).
Recall that smoothness was one condition on the momentum space
kernels (\ref{3.5}) in our renormalization problem. Therefore,
we will assume that $G\in{\mathcal C}^\infty(\R^N)$ and look
for solutions $F\in{\mathcal C}^\infty(\R^N)$ of (\ref{5.1}).
The right hand side of (4.6) is a sum of multiple convolutions
of lower order vertices with cutoff propagators. With an
exponential cutoff, smoothness therefore iterates through the
induction. But it is luxury in the sense that no higher momentum
space derivatives are needed than those required by the Taylor
expansions of the non--irrelevant momentum space kernels.
The reader is invited to diminish the demands on regularity
to this situation in the following statements.

\subsection{Irrelevant case}

Consider first the case when
$\sigma <0$. We can then immediately integrate (\ref{5.1}). It
is equivalent to
\begin{equation}
L\dtota{L}\left\{L^{-\sigma}F(Lp)\right\}=
L^{-\sigma}G(Lp).
\label{5.2}
\end{equation}
A special solution thereto is
\begin{equation}
F(p)=\int_0^1\dscale{L}\;L^{-\sigma}G(Lp).
\label{5.3}
\end{equation}
The integral converges for $\sigma <0$ and yields a smooth
function $F(p)$. Furthermore, $F(p)$ inherits the symmetries
of $G(p)$.

The integral (\ref{5.3}) exhausts the irrelevant case. Suppose
that $F_1(p)$ and $F_2(p)$ are two different smooth solutions to
(\ref{5.2}). Their difference $\bigtriangleup F(p)=F_2(p)-F_1(p)$
obeys the homogeneous equation
\begin{equation}
\left\{p\dpart{p}-\sigma\right\}\bigtriangleup F(p)=0,
\label{5.4}
\end{equation}
and is consequently a homogeneous function in $p$ of degree
$\sigma$. But regularity at $p=0$ demands $\bigtriangleup F(p)=0$,
when $\sigma <0$.
\\[4mm]{\sl
Let $\sigma <0$ and $G\in{\mathcal C}^\infty(\R^N)$. Then we
have that: (1) There exists a unique solution $F(p)$ of the PDE
(\ref{5.1}) in the space ${\mathcal C}^\infty(\R^N)$. (2) It is
given by the integral (\ref{5.3}).
}\\[2mm]
\subsection{Non--irrelevant case}

Consider then the case
$\sigma\geq 0$. The integral (\ref{5.3}) now fails to converge
unless $G(p)$ provides a sufficiently high power of $p$. This
can be achieved by a Taylor expansion of order $\sigma$. The
remainder term behaves as $O(L^{\sigma +1})$ and compensates
the power $L^{-1-\sigma}$ in (\ref{5.3}). We thus expand
\begin{equation}
F(p)=\sum_{|\alpha|\leq\sigma}\frac{p^\alpha}{\alpha!}
\frac{\partial^{|\alpha|}F}{\partial p^\alpha}(0)+
\sum_{|\alpha|=\sigma+1}\frac{p^\alpha}{\alpha!}
\int_0^1{\mathrm d}t\;(1-t)^{|\alpha|-1}
\frac{\partial^{|\alpha|}F}{\partial p^\alpha}(tp),
\label{5.5}
\end{equation}
where $\alpha=(\alpha_1,\ldots,\alpha_N)\in\N^N$ is an integer
valued multi--index, $|\alpha|=\alpha_1+\cdots+\alpha_N$,
$\alpha!=\alpha_1!\cdots\alpha_N!$, and $p^\alpha=
p_1^{\alpha_1}\cdots p_N^{\alpha_N}$. Taking derivatives of
(\ref{5.1}), we find that
\begin{equation}
\left\{p\dpart{p}-(\sigma-|\alpha|)\right\}
\frac{\partial^{|\alpha|}F}{\partial p^\alpha}(p)=
\frac{\partial^{|\alpha|}G}{\partial p^\alpha}(p).
\label{5.6}
\end{equation}
Each $p$--derivative thus reduces $\sigma$ by one unit. The
derivatives of order $\sigma+1$ in (\ref{5.5}) therefore fall
into the irrelevant case. From (\ref{5.3}) it follows that
\begin{equation}
\frac{\partial^{|\alpha|}F}{\partial p^\alpha}(p)=
\int_0^1\dscale{L}\;L^{-(\sigma-|\alpha|)}
\frac{\partial^{|\alpha|}G}{\partial p^\alpha}(Lp),
\label{5.7}
\end{equation}
for all derivatives of order $|\alpha|=\sigma+1$ (in fact
$|\alpha|\geq\sigma+1$). The remainder term in (\ref{5.5})
then involves a convergent integral. It yields a smooth
function. The Taylor coefficients in (\ref{5.5}) on the
other hand are determined by evaluation of (\ref{5.6}) at
$p=0$. We have that
\begin{equation}
-(\sigma-|\alpha|)
\frac{\partial^{|\alpha|}F}{\partial p^\alpha}(0)=
\frac{\partial^{|\alpha|}G}{\partial p^\alpha}(0),
\label{5.8}
\end{equation}
for all solutions to (\ref{5.6}) which are
regular at $p=0$. Eq. (\ref{5.8}) determines all relevant
Taylor coefficients with $|\alpha|<\sigma$ (in fact all
non--marginal ones with $|\alpha|\neq\sigma$). The marginal
Taylor coefficients are not only undetermined by (\ref{5.8})
but also impose the condition
\begin{equation}
\frac{\partial^{|\alpha|}G}{\partial p^\alpha}(0)=0
\label{5.9}
\end{equation}
on the Taylor coefficients of $G(p)$ of order $\sigma$. Smooth
solutions exist only if the PDE (\ref{5.6}) satisfies
(\ref{5.9}). If this is the case, as will be supposed here,
then the Taylor coefficients of $F(p)$ of order
$\sigma$ are free parameters. In the perturbation theory for
the $\phi^4$--trajectory, we will use this freedom of
the marginals at a given order of perturbation theory to
satisfy the conditions on the existence of smooth solutions
one order below. In this sense, the condition (\ref{5.9})
on the marginal Taylor coefficients determines both the
$\beta$--function and the wave function renormalization.
\\[4mm]{\sl
Let $\sigma\geq 0$ and $G\in{\mathcal C}^\infty(\R^N)$. Then
we have that: (1) The PDE (\ref{5.1}) has smooth solutions
$F(p)$ if and only if all derivatives of $G(p)$ of order
$\sigma$ satisfy (\ref{5.9}). Let this be the case. (2) Then
$F(p)$ can be written in the form of a Taylor expansion
(\ref{5.5}) of order $\sigma$. The Taylor remainder follows
from (\ref{5.7}). The Taylor coefficients of order $\sigma$
are free parameters. The Taylor coefficients of order less
than $\sigma$ follow from (\ref{5.8}).
}\\[4mm]
There is a neat reformulation for the non--irrelevant case in
terms of subtracted functions. Eq. (\ref{5.1}) is equivalent
to
\begin{equation}
\left\{L\dtota{L}-\sigma\right\}F(Lp)=G(Lp).
\label{5.10}
\end{equation}
Smooth solutions to (\ref{5.10}) also obey
\begin{equation}
\left\{L\dtota{L}-(\sigma -n) \right\}
\frac{d^n}{dL^n}F(Lp)=
\frac{d^n}{dL^n}G(Lp),
\label{5.11}
\end{equation}
for all $n\geq 0$. In particular, we have the identities
\begin{equation}
-(\sigma -n)\frac{d^n}{dL^n}F(Lp)\biggl\vert_{L=0}=
\frac{d^n}{dL^n}G(Lp)\biggl\vert_{L=0}
\label{5.12}
\end{equation}
as polynomials in $p$. We can then define subtracted functions
\begin{equation}
F^{(n)}(Lp)=F(Lp)-\sum_{m=0}^n\frac{L^m}{m!}
\frac{d^m}{dL^m}F(Lp)\biggr\vert_{L=0}.
\end{equation}
From (\ref{5.10}) and (\ref{5.12}) it follows that they
satisfy the subtracted version of (\ref{5.10}), namely
\begin{equation}
\left\{L\dtota{L}-\sigma\right\}F^{(n)}(Lp)=
G^{(n)}(Lp).
\label{5.14}
\end{equation}
But the subtracted function has the property
$G^{(n)}(Lp)=O(L^{n+1})$ for all $p\in\R^N$. Therefore,
the integral
\begin{equation}
F^{(n)}(p)=\int_0^1\dscale{L}\;L^{-\sigma}
G^{(n)}(Lp)
\label{5.15}
\end{equation}
converges for $n\geq\sigma$. The Taylor remainder can
thus be integrated as in the irrelevant case.

\section{Perturbation theory}

Renormalized perturbation theory is the inductive solution of
the scaling equations (\ref{4.6}). We have included a separate
treatment of the first and second order scaling equations to
illustrate the method.

\subsection{First order}

In the functional representation,
the scaling equation (\ref{4.1}) requires for $r=1$ that
\begin{equation}
\left\{\pertb{1}-\dilat{\phi}\right\}
\pertv{1}{\phi}=0.
\label{6.1}
\end{equation}
In other words, $\pertv{1}{\phi}$ has to be an eigenvector of
$\left({\mathcal D}\phi,\delta_\phi\right)$, and $\pertb{1}$
has to be the eigenvalue. The $\phi^4$--vertex
\begin{equation}
\obs{4,0}{\phi}=\frac{1}{4!}\int{\mathrm d}^Dx\;\phi(x)^4
\label{6.2}
\end{equation}
is an eigenvector of $\left({\mathcal D}\phi,\delta_\phi\right)$.
Its eigenvalue is $\pertb{1}=4-D$. In four dimensions, the
$\phi^4$--vertex becomes marginal. The wave function term
\begin{equation}
\obs{2,2}{\phi}=\frac{1}{2}\int{\mathrm d}^Dx\;
\phi(x)(-\bigtriangleup )\phi(x)
\label{6.3}
\end{equation}
is marginal in any dimension. We therefore conclude that
\begin{equation}
\pertv{1}{\phi}=
\obs{4,0}{\phi}+\pertz{1}\obs{2,2}{\phi}
\label{6.4}
\end{equation}
is in agreement with (\ref{6.1}), whence $\pertb{1}=0$. The
first order wave function parameter $\pertz{1}$ is not
determined by (\ref{6.1}).

\subsection{Second order}

The first order equation (\ref{6.1})
is special as it is homogeneous. All higher order equations are
inhomogeneous. To second order, (\ref{4.1}) reads
\begin{equation}
\left\{2\pertb{1}-\dilat{\phi}\right\}
\pertv{2}{\phi}=-\pertk{2}{\phi},
\label{6.5}
\end{equation}
where
\begin{equation}
\pertk{2}{\phi}=\pertb{2}\pertv{1}{\phi}+
2\bilin{\pertv{1}{\phi}}{\pertv{1}{\phi}}.
\label{6.6}
\end{equation}
Eq. (\ref{6.6}) contains two unknowns, $\pertb{2}$ and $\pertz{1}$.
They belong to the two marginal couplings. In the momentum space
kernel representation, (\ref{6.5})  becomes
\begin{equation}
\left\{\pertsp{2}{2n}-\sum_{m=1}^{2n}p_m\dpart{p_m}\right\}
\pkern{2}{2n}=\pkker{2}{2n},
\label{6.7}
\end{equation}
with
\begin{eqnarray}
\widetilde{{\mathcal K}}^{(2)}_2(p_1,p_2)&=&
\pertz{1}A+\pertb{2}\pertz{1}p_1^2+
2(\pertz{1}p_1^2)^2\widetilde{C}(p_1)+
\nonumber\\&&\quad
\widetilde{C}\star\widetilde{v}\star\widetilde{v}(p_1),
\label{6.8}\\
\pkker{2}{4}&=&
\pertb{2}+8\pertz{1}p_1^2\widetilde{C}(p_1)+
6\widetilde{C}\star\widetilde{v}(p_1+p_2),
\label{6.9}\\
\pkker{2}{6}&=&
20\widetilde{C}(p_1+p_2+p_3),
\label{6.10}
\end{eqnarray}
where $\star$ denotes convolution in momentum space times
$(2\pi)^{-D}$. The kernels are understood to be symmetrized
in their entries and to be restricted to the hyperplane of
zero total momentum. Put $p_{2n}=-\sum_{m=1}^{2n-1}p_{m}$
for instance. The constant $A$ in (\ref{6.8}) stands for the
convergent one loop integral
\begin{equation}
A=4\int\dmom{D}{p}\;e^{-2p^2}=\frac{1}{(4\pi)^2},\quad D=4.
\label{6.11}
\end{equation}
All higher kernels with $n>3$ are zero in accordance with
(\ref{3.4}). The six point kernel has scaling dimension
$\pertsp{r}{6}=-2$ and is irrelevant. From (\ref{5.3}) we
learn that it is given by\footnote{It is instructive to
perform this integral. The result
is $\wt{\VV}^{(2)}_6(p_1,\ldots,p_5)=10(p_1+p_2+p_3)^{-2}
\left\{\exp\left(-(p_1+p_2+p_3)^2\right)-1\right\}$. This
expression is regular at zero momentum and a bounded
function of the momenta. It has the form of a cutoff
propagator.}
\begin{equation}
\pkern{2}{6}=-\int_0^1\dscale{L}\;L^2
\widetilde{{\mathcal K}}^{(2)}_6(Lp_1,\ldots,Lp_6).
\label{6.12}
\end{equation}
Notice that (\ref{6.10}) is indeed a smooth function of
the momenta. The four point kernel has scaling dimension
$\pertsp{r}{4}=0$ and is marginal. It requires a
separate treatment of its zero momentum value, the four
point coupling. We determine $\pertb{2}$ such that
\begin{equation}
\wt{{\mathcal K}}^{(2)}_{4}(0,0,0,0)=0.
\label{6.13}
\end{equation}
From (\ref{6.9}) it follows that the condition (\ref{6.13})
reads
\begin{equation}
\pertb{2}=-6\widetilde{C}\star\widetilde{v}(0)=
\frac{-6}{(4\pi)^2},\quad D=4.
\label{6.14}
\end{equation}
Notice that it is independent of $\pertz{1}$. The value of
the second order $\phi^4$--coupling is not determined by
renormalization invariance but rather by the choice of the
expansion parameter. A natural choice is (\ref{3.8}). The
four point kernel is then equal to its first subtraction.
It is therefore integrated to
\begin{equation}
\pkern{2}{4}=-\int_0^1\dscale{L}\;
\widetilde{{\mathcal K}}^{(2)}_4(Lp_1,\ldots,Lp_4).
\label{6.15}
\end{equation}
The integral converges due to (\ref{6.13}). The flow of the
coupling parameter thus saves us from a logarithmic
singularity of (\ref{6.15}). We remark that, in three
dimensions, the four point kernel is already irrelevant
to second order and needs no subtraction at all. The
quadratic kernel finally has scaling dimension
$\pertsp{r}{2}=2$ and is relevant. It calls for a Taylor
expansion of order two with remainder term. Using
Euclidean invariance, we write
\begin{eqnarray}
\widetilde{{\mathcal V}}^{(2)}_2(p,-p)&=&F(p^2),
\label{6.16}\\
\widetilde{{\mathcal K}}^{(2)}_2(p,-p)&=&2G(p^2).
\label{6.17}
\end{eqnarray}
Furthermore, we trade $p^2$ for a new variable $u$.
The scaling equation for the quadratic kernel then
takes the form
\begin{equation}
\left\{1-u\dtota{u}\right\}F(u)=G(u).
\label{6.18}
\end{equation}
To solve it, we expand $F(u)$ in a Taylor formula
\begin{equation}
F(u)=F(0)+F^{\prime}(0)u+
\frac{u^2}{2}\int_{0}^{1}{\mathrm d}t\;(1-t)\;
F^{\prime\prime}(tu).
\label{6.19}
\end{equation}
The Taylor coefficients follow from evaluating (\ref{6.18}) and
its $u$--derivatives at $u=0$. We find
\begin{equation}
F(0)=G(0)=
\frac{1}{2}\widetilde{C}\star\widetilde{v}\star\widetilde{v}(0)+
\pertz{1}\frac{A}{2},
\label{6.20}
\end{equation}
with a convergent two loop integral, which comes out as
\begin{equation}
\frac{1}{2}\widetilde{C}\star\widetilde{v}\star\widetilde{v}(0)=
\frac{1}{(4\pi)^{2}}\left(2\log (2)-\log (3)\right).
\label{6.21}
\end{equation}
The other Taylor coefficient is the second order wave function
parameter
\begin{equation}
F^\prime (0)=\pertz{2}.
\label{6.22}
\end{equation}
We leave it undetermined for the moment. Taking a $u$-derivative
of (\ref{6.18}), it follows that
\begin{equation}
-u\dtota{u}F^\prime(u)=G^\prime(u).
\label{6.23}
\end{equation}
Regularity of $F^\prime(u)$ at $u=0$ requires that
\begin{equation}
0=2\;G^\prime(u)=\pertb{2}\pertz{1}+\dpart{p^2}
\widetilde{C}\star\widetilde{v}\star\widetilde{v}(p)
\biggr\vert_{p=0}.
\label{6.24}
\end{equation}
This condition determines the first order wave function parameter.
Its value in four dimensions is
\begin{equation}
\pertz{1}=\frac{-1}{18(4\pi)^2}.
\label{6.25}
\end{equation}
Eq. (\ref{6.24}) explains the presence of a wave function term in
(\ref{3.3}). It is necessary to fulfill the condition under which
smooth solutions of the second order equation (\ref{6.18}) exist.
The second order wave function parameter (\ref{6.22}) is then
fixed by the condition on the third order pendant to (\ref{6.24}).
The value of the second order mass parameter (\ref{6.20}) is
then
\begin{equation}
F(0)=
\frac{1}{(4\pi)^2}\left(2\log (2)-\log (3)-\frac{1}{36}\right).
\label{6.26}
\end{equation}
A second $u$--derivative turns (\ref{6.23}) into
\begin{equation}
\left\{-1-u\dtota{u}\right\}F^{\prime\prime}(u)=
G^{\prime\prime}(u).
\label{6.27}
\end{equation}
The second derivative is thus irrelevant. Eq. (\ref{6.27})
is equivalent to
\begin{equation}
t\dtota{t}\left\{t\;F^{\prime\prime}(tu)\right\}=
-t\;G^{\prime\prime}(tu).
\label{6.28}
\end{equation}
Therefrom we conclude that
\begin{equation}
F^{\prime\prime}(u)=
-\int_0^1\dscale{t}\;t\;G^{\prime\prime}(tu).
\label{6.29}
\end{equation}
The integrals (\ref{6.19}) and (\ref{6.29}) will not be evaluated.
We content ourselves with the ascertainment that they converge and
yield a smooth remainder term. The second order scheme is now
complete.

\subsection{Higher orders}

The second order scheme generalizes
to all higher orders. To order $r$, we first compute $\pertb{r}$,
then $\pertz{r-1}$, and thereafter $\pertv{r}{\phi}$, except for
$\pertz{r}$. Then we proceed to the order $r+1$. Each step consists
of solving a system of scaling equations (\ref{4.6}) for the
order $r$ momentum space kernels.

\subsubsection{Taylor coefficients}

The momentum space kernels with non--negative scaling dimension
will have to be Taylor expanded. In four dimensions, the quadratic
needs to be expanded to second order, whereas the quartic kernel
needs to be expanded to first order. A convenient way of writing
the Taylor expansions is in terms of
projectors
\begin{eqnarray}
\pro{2,0}\pkern{r}{2n}&=&\delta_{n,1}
\widetilde{{\mathcal V}}^{(r)}_2(0,0),
\label{6.30}\\
\pro{2,2}\pkern{r}{2n}&=&\delta_{n,1}
\dpart{(p^2)}\widetilde{{\mathcal V}}^{(r)}_2(p,-p)\bigr\vert_{p=0},
\label{6.31}\\
\pro{4,0}\pkern{r}{2n}&=&\delta_{n,2}
\widetilde{{\mathcal V}}^{(r)}_{4}(0,0,0,0).
\label{6.32}
\end{eqnarray}
They are defined to act linearly on polynomial functionals
of the form (\ref{3.4}). We have that
\begin{eqnarray}
\pro{2,0}\pertv{r}{\phi}&=&\mu^{(r)}\obs{2,0}{\phi},
\label{6.33}\\
\pro{2,2}\pertv{r}{\phi}&=&\pertz{r}\obs{2,2}{\phi},
\label{6.34}\\
\pro{4,0}\pertv{r}{\phi}&=&\lambda^{(r)}\obs{2,0}{\phi},
\label{6.35}
\end{eqnarray}
with coupling parameters
\begin{eqnarray}
\mu^{(r)}&=&\widetilde{{\mathcal V}}^{(r)}_2(0,0),
\label{6.36}\\
\pertz{r}&=&
\dpart{(p^2)}\widetilde{{\mathcal V}}^{(r)}_2(p,-p)\bigr\vert_{p=0},
\label{6.37}\\
\lambda^{(r)}&=&\widetilde{{\mathcal V}}^{(r)}_{4}(0,0,0,0),
\label{6.38}
\end{eqnarray}
and local vertices
\begin{eqnarray}
\obs{2,0}{\phi}&=&\frac{1}{2}\int{\mathrm d}^Dx\;\phi(x)^2,
\label{6.39}\\
\obs{2,2}{\phi}&=&\frac{1}{2}\int{\mathrm d}^Dx\;
\phi(x)(-\bigtriangleup)\phi(x),
\label{6.40}\\
\obs{4,0}{\phi}&=&\frac{1}{4!}\int{\mathrm d}^Dx\;\phi(x)^4.
\label{6.41}
\end{eqnarray}
They are all eigenvectors of $\left({\mathcal D}\phi,
\delta_{\phi}\right)$ with eigenvalues
\begin{eqnarray}
\dilat{\phi}\obs{2,0}{\phi}&=&2\;\obs{2,0}{\phi},
\label{6.42}\\
\dilat{\phi}\obs{2,2}{\phi}&=&0,
\label{6.43}\\
\dilat{\phi}\obs{4,0}{\phi}&=&(4-D)\obs{4,0}{\phi}.
\label{6.44}
\end{eqnarray}
When applied to (\ref{4.1}), the projectors yield the
scaling equations
\begin{eqnarray}
\left(r\pertb{1}-2\right)\mu^{(r)}\obs{2,0}{\phi}&=&
-\pro{2,0}{\mathcal K}^{(r)}(\phi),
\label{6.45}\\
r\pertb{1}\pertz{r}\obs{2,2}{\phi}&=&
-\pro{2,2}{\mathcal K}^{(r)}(\phi),
\label{6.46}\\
\left(r\pertb{1}-(4-D)\right)\lambda^{(r)}\obs{4,0}{\phi}&=&
-\pro{4,0}{\mathcal K}^{(r)}(\phi).
\label{6.47}
\end{eqnarray}
By means of (\ref{4.2}), these are given by
\begin{eqnarray}
\pro{2,0}{\mathcal K}^{(r)}(\phi)&=&
\sum_{s=2}^{r}\binox{r}{s}\pertb{s}\mu^{(r-s+1)}
\obs{2,0}{\phi}\nonumber\\& &
+\sum_{s=1}^{r-1}\binox{r}{s}\pro{2,0}
\bilin{\pertv{s}{\phi}}{\pertv{r-s}{\phi}},
\label{6.48}\\
\pro{2,2}{\mathcal K}^{(r)}(\phi)&=&
\sum_{s=2}^{r}\binox{r}{s}\pertb{s}\pertz{r-s+1}
\obs{2,2}{\phi}\nonumber\\& &
+\sum_{s=1}^{r-1}\binox{r}{s}\pro{2,2}
\bilin{\pertv{s}{\phi}}{\pertv{r-s}{\phi}},
\label{6.49}\\
\pro{4,0}{\mathcal K}^{(r)}(\phi)&=&
\sum_{s=2}^{r}\binox{r}{s}\pertb{s}\lambda^{(r-s+1)}
\obs{4,0}{\phi}\nonumber\\& &
+\sum_{s=1}^{r-1}\binox{r}{s}\pro{4,0}
\bilin{\pertv{s}{\phi}}{\pertv{r-s}{\phi}}.
\label{6.50}
\end{eqnarray}
Eqs. (\ref{6.45}), (\ref{6.46}), and (\ref{6.47}) are linear
equations for the coupling parameters (\ref{6.36}), (\ref{6.37}),
and (\ref{6.38}). They are algebraic equations rather than
first order differential equations.

\subsubsection{The coefficient $\pertb{r}$}

In four dimensions, (\ref{6.47}) reads
\begin{equation}
\pro{4,0}{\mathcal K}^{(r)}(\phi)=
{\mathcal K}^{(r)}_4(0,0,0,0)\;\obs{4,0}{\phi}=0,
\label{6.51}
\end{equation}
since $\pertb{1}=0$. The choice (\ref{3.8}) of the coupling
parameter means $\lambda^{(s)}=\delta_{s,1}$. From
(\ref{6.50}) we then have
\begin{equation}
\pertb{r}\obs{4,0}{\phi}=
-\sum_{s=1}^{r}\binox{r}{s}\pro{4,0}
\bilin{\pertv{s}{\phi}}{\pertv{r-s}{\phi}}.
\label{6.52}
\end{equation}
Its right hand side is known from the induction hypothesis,
and does not depend on $\pertz{r-1}$ because
\begin{eqnarray}
\pro{4,0}\bilin{\obs{4,0}{\phi}}{\obs{2,2}{\phi}}&=&0,
\label{6.53}\\
\pro{4,0}\bilin{\obs{2,2}{\phi}}{\obs{2,2}{\phi}}&=&0,
\label{6.54}
\end{eqnarray}
due to the regularity of $\widetilde{C}(p)$ at $p=0$. Eq.
(\ref{6.52}) determines $\pertb{r}$.

\subsubsection{The coefficient $\pertz{r}$}

Consider then (\ref{6.46}). Since $\pertb{1}=0$, it
becomes
\begin{equation}
\pro{2,2}{\mathcal K}^{(r)}(\phi)=
\dpart{(p^2)}{\mathcal K}^{(r)}_2(p,-p)\biggr\vert_{p=0}
\obs{2,2}{\phi}=0.
\label{6.55}
\end{equation}
From (\ref{6.49}) it follows that
\begin{eqnarray}
\binox{r}{2}\pertb{2}\pertz{r-1}\obs{2,2}{\phi}&=&
-\sum_{s=3}^{r}\binox{r}{s}\pertb{s}\pertz{r-s+1}\obs{2,2}{\phi}
\nonumber\\& &-\sum_{s=1}^{r-1}\binox{r}{s}
\pro{2,2}\bilin{\pertv{s}{\phi}}{\pertv{r-s}{\phi}}.
\label{6.56}
\end{eqnarray}
The right hand side is again independent of $\pertz{r-1}$
because
\begin{eqnarray}
\pro{2,2}\bilin{\obs{4,0}{\phi}}{\obs{2,2}{\phi}}&=&0,
\label{6.57}\\
\pro{2,2}\bilin{\obs{2,2}{\phi}}{\obs{2,2}{\phi}}&=&0.
\label{6.58}
\end{eqnarray}
But then (\ref{6.35}) determines $\pertz{r-1}$.

\subsubsection{The coefficient $\mu^{(r)}$}

The last coefficient is the effective mass parameter.
Eq. (\ref{6.45}) tells that
\begin{equation}
2\mu^{(r)}\obs{2,0}{\phi}=
\pro{2,0}\pertk{r}{\phi},
\label{6.59}
\end{equation}
and using (\ref{6.48}) we then have
\begin{eqnarray}
2\mu^{(r)}\obs{2,0}{\phi}&=&
\sum_{s=2}^{r}\binox{r}{s}\pertb{s}
\mu^{(r-s+1)}\obs{2,0}{\phi}
\nonumber\\
& &+\sum_{s=1}^{r-1}\binox{r}{s}
\pro{2,0}\bilin{\pertv{s}{\phi}}{\pertv{r-s}{\phi}}.
\label{6.60}
\end{eqnarray}
The non--irrelevant part of $\pertv{r}{\phi}$ is now complete,
except for the wave function parameter $\pertz{r}$ which we
leave undetermined until the next order.

\subsubsection{Irrelevant part}

The irrelevant part is directly integrated. For $n\geq 3$, the
momentum space scaling dimension (\ref{4.8}) is negative. Those
kernels are therefore all irrelevant. The scaling equation
(\ref{4.6}) is in this case integrated to
\begin{equation}
\pkern{r}{2n}=-\int_0^1\dscale{L}\;L^{-\pertsp{r}{2n}}
\widetilde{{\mathcal K}}^{(r)}_{2n}(Lp_1,\ldots,Lp_{2n}).
\label{6.61}
\end{equation}
All kernels with $n>r+1$ are zero. The integrals converge due
to the negative power counting.

The zero momentum value of the four point kernel has been
transfered to the differential $\beta$-function through
(\ref{6.51}). Its
subtracted remainder is irrelevant and integrated to
\begin{equation}
\pkern{r}{4}=-\int_0^1\dscale{L}\;
\widetilde{{\mathcal K}}^{(r)}_4(Lp_1,\ldots,Lp_4).
\label{6.62}
\end{equation}
The integral converges due to the subtraction at zero momentum,

Finally, the two point kernel is reconstructed with the help of
\begin{eqnarray}
\widetilde{{\mathcal V}}^{(r)}_2(p,-p)&=&F(p^2),
\label{6.63}\\
\widetilde{{\mathcal K}}^{(r)}_2(p,-p)&=&2G(p^2),
\label{6.64}
\end{eqnarray}
and
\begin{equation}
F(u)=\mu^{(r)}+\zeta^{(r)}u+
\frac{u^2}{2}\int_0^1{\rm d}s(1-s)F^{\prime\prime}(su)
\label{6.65}
\end{equation}
through
\begin{equation}
G^{\prime\prime}(u)=
-\int_0^1 \frac{{\rm d}t}{t}\;t G^{\prime\prime}(tu).
\label{6.66}
\end{equation}
The iterative scheme is now complete, aside of the question of
the convergence and smoothness of the right hand side of
(\ref{4.6}).

\section{Bilinear renormalization group bracket}

The bilinear renormalization group bracket is responsible for
inhomogeneous terms in our scaling equations. We write it out
explicitly for even monomials $\obs{2n}{\phi}$ of the general
form
\begin{equation}
\obs{2n}{\phi}=
\frac 1{(2n)!}\intex{2n}\;\phixs{2n}
\;\obs{2n}{x_1,\dots,x_{2n}}.
\label{7.1}
\end{equation}
Applied to two monomials of this form (\ref{7.1}), the
bilinear bracket decomposes into
\begin{equation}
\left\langle \obs{2n}{\phi},\obs{2m}{\phi}\right\rangle=
\sum_{l=|n-m|}^{n+m-1} N_{n,m,l}
\left({\mathcal O}_{2n}\star{\mathcal O}_{2m}\right)_{2l}(\phi),
\label{7.2}
\end{equation}
and is itself a sum of monomials
\begin{gather}
\left({\mathcal O}_{2n}\star{\mathcal O}_{2m}\right)_{2l}(\phi)=
\nonumber \\
\frac 1{(2l)!}\intex{2l}\;\phixs{2l}\;
\left({\mathcal O}_{2n}\star{\mathcal O}_{2m}\right)_{2l}
(x_1,\dots,x_{2l}),
\label{7.3}
\end{gather}
whose kernels are given by a multiple convolutions
\begin{eqnarray}
&&\left({\mathcal O}_{2n}\star{\mathcal O}_{2m}\right)_{2l}
(x_1,\dots,x_{2l})=
\nonumber\\&&\quad
\frac 1{2(2l)!}
\int{\rm d}y_1\dots{\rm d}y_{2(n+m-l)}
\;C(y_1-y_{n+m-l+1}) \prod_{k=2}^{n+m-l} v(y_k-y_{n+m-l+k})
\nonumber\\&&\quad
\biggl\{\obs{2n}{(x_1,\dots,x_{n-m+l},y_1,\dots,y_{n+m-l})}
\nonumber\\&&\quad\quad
\obs{2m}{(x_{n-m+l+1},\dots,x_{2l},y_{n+m-l+1},\dots,
y_{2(n+m-l)})}+
\nonumber\\&&\quad\quad
((2l)!-1)\; {\rm permutations}\biggr\}.
\label{7.4}
\end{eqnarray}
The kernels are understood to be symmetric under permutations
of their entries. The multiple convolution in (\ref{7.4})
always involves one hard propagator $C$ and $n+m-l-1$ soft
propagators $v$, which is at the same time the number of
loops. Furthermore, (\ref{7.2}) involves a combinatorial
factor
\begin{equation}
N_{n,m,l}=
\frac{(2l)!}
{(n+m-l-1)!(n-m+l)!(m-n+l)!},
\label{7.5}
\end{equation}
coming from the number of ways in which the contractions
can be made. Eq. (\ref{7.4}) can be interpreted as a fusion
of two vertices. In the process of fusion links are created,
consisting of propagators.

We present an elementary estimate on this fusion product.
The estimate uses the $L_{\infty,\epsilon}$-norm in momentum
space. The estimate works at this point for $\epsilon \geq 0$.
Later it will be used for $\epsilon >0$ only. Notice to begin
with that
\begin{equation}
\| \wt{C}\|_{\infty,-2\epsilon}<\infty, \quad
\| \wt{v}\|_{1,-2\epsilon}<\infty,
\label{7.6}
\end{equation}
for the propagators with exponential cutoff.\footnote{Here
$\| \wt{C}\|_{\infty,-2\epsilon}={\sup }_{p\in\R^D}
\{|\wt{v}(p)| e^{2\epsilon |p|}\} $ and
$\| \wt{v}\|_{1,-2\epsilon} =(2\pi )^{-D}\int{\rm d}^Dp
|\wt{v}(p) |e^{2\epsilon |p|}$
denote the $L_{\infty,-2\epsilon}$- and
$L_{1,-2\epsilon}$-norms in momentum space.}
At $\epsilon =0$ we have for instance $\|\wt{C}\|_\infty=2$ and
$\|\wt{v}\|_\infty\leq 2\pi^{D/2}/(D-2)$. If the Fourier
transformed kernels now satisfy the bounds
\begin{equation}
\| \wt{{\mathcal O}}_{2n}\|_{\infty,\epsilon}<\infty,\quad
\| \wt{{\mathcal O}}_{2m}\|_{\infty,\epsilon} < \infty,
\label{7.7}
\end{equation}
that is, are finite in the
$L_{\infty,\epsilon}$-norm,\footnote{The
$L_{\infty,\epsilon}$-norm for the momentum space kernels is
defined as $\| \wt{{\mathcal O}}_{2n}\|_{\infty,\epsilon} =
{\rm sup}_{(p_1,\ldots,p_{2n})
\in {\cal P}_{2n}}\{|\wt{{\mathcal O}}(p_1,\ldots,p_{2n})|
e^{-\epsilon (|p_1|+\cdots +|p_{2n}|)}\}$ with
${\cal P}_{2n}=\{ (p_1,\ldots ,p_{2n})\in \R^D \times\cdots
\times\R^D | p_1+\cdots +p_{2n}=0\}$ the hyperplane of
total zero momentum.} then it follows that all summands
(\ref{7.4}) in the decomposition (\ref{7.2}) of the bilinear
operation have finite $L_{\infty,\epsilon}$-norms in momentum
space. They obey
\begin{equation}
\| ({\mathcal O}_{2n}\star
{\mathcal O}_{2m})^\sim_{2l}\|_{\infty,\epsilon}
\leq \frac{1}{2}
\;\| \wt{C}\|_{\infty,-2\epsilon}
\;\| \wt{v}\|_{1,-2\epsilon}^{n+m-l-1}
\;\|\wt{{\mathcal O}}_{2n}\|_{\infty,\epsilon}
\;\|\wt{{\mathcal O}}_{2m}\|_{\infty,\epsilon}.
\label{7.8}
\end{equation}
Therefore, the renormalization group flow preserves the
$L_{\infty,\epsilon}$-norm of momentum space kernels.
The estimate immediately follows from the Fourier transform
\begin{eqnarray}
&&\left({\mathcal O}_{2n}\star{\mathcal O}_{2m}\right)^\sim_{2l}
(p_1,\dots,p_{2l})=
\nonumber\\&&\quad
\frac 1{2(2l)!}
\int\frac{{\rm d}^Dq_1}{(2\pi)^D}\dots
\frac{{\rm d}^Dq_{n+m-l-1}}{(2\pi)^D}
\;\wt{C}(q_{n+m-l})
\;\prod_{k=1}^{n+m-l-1} \wt{v}(q_i)
\nonumber\\&&\quad
\biggl\{\wt{{\mathcal O}}_{2n}(p_1,\dots,p_{n-m+l},
q_{1},\dots,q_{n+m-l})
\nonumber\\&&\quad\quad
\wt{{\mathcal O}}_{2m}(p_{n-m+l+1},\dots,p_{2l},-q_{1},\dots,
-q_{n+m-l})+
\nonumber\\&&\quad\quad
((2l)!-1)\; {\rm permutations}\biggr\}.
\label{7.9}
\end{eqnarray}
of (\ref{7.4}). The $\delta$-functions from translation
invariance have been removed. The sums of momenta in
the kernels are zero through
\begin{equation}
p_{2l}=-\sum_{m=1}^{2l-1}p_m,\quad
q_{n+m-l}=-\sum_{k=1}^{n-m+l}p_k-\sum_{k=1}^{n+m-l-1}q_k.
\label{7.10}
\end{equation}
The idea with the parameter $\epsilon$ is to use part
of the exponential large momentum decay of the fluctuation
and normal ordering propagators to compensate a possible large
momentum growth of the kernels.

In an initial value problem for a scale dependent
renormalization group flow with $L_\infty$--bounded initial data
this might seem unnecessary. For instance, a pure $\phi^4$--vertex
is given by a constant kernel and is thus $L_\infty$--bounded. The
evolution preserves the $L_\infty$--bound for all {\it finite}
scales $L$. However, we cannot expect the solution to be
$L_\infty$--bounded {\it uniformly} in $L$. The limit
$L\rightarrow\infty$ requires a separate treatment of zero momentum
derivatives and Taylor remainders of the non-irrelevant kernels.
The price to pay is a growth in momentum space. In the four
dimensional case it is polynomial in powers and logarithms of
momenta. With an exponential bound we are very far on
the safe side.

\section{Large momentum bound}

We prove a large momentum bound for the solution to the
renormalization group PDE (\ref{5.1}) under the assumption
of a large momentum bound on the inhomogeneous side.

The bound uses the $L_{\infty,\epsilon}$--norm for
$\epsilon >0$. Although being rather wasteful, it suffices
to prove finiteness of the bilinear renormalization group
bracket.

\subsection{Irrelevant case}

Let $\sigma <0$. Suppose that the function $G(p)$ in
(\ref{5.1}) has a finite $L_{\infty,\epsilon}$--norm
\begin{equation}
\| G\|_{\infty,\epsilon}=
\sup_{p\in\R ^N}
\left\{ |G(p)| e^{-\epsilon |p|} \right\}<\infty.
\label{8.1}
\end{equation}
Its solution (\ref{5.3}) then inherits an
$L_{\infty,\epsilon}$--bound. From
\begin{equation}
|F(p)| e^{-\epsilon |p|}\leq
\int_0^1 \frac{{\rm d}L}{L} L^{-\sigma}
|G(Lp)| e^{-\epsilon |p|}
\leq
\int_0^1 \frac{{\rm d}L}{L} L^{-\sigma}
e^{-(1-L)\epsilon |p|}
\| G\|_{\infty,\epsilon}
\label{8.2}
\end{equation}
it follows that
\begin{equation}
\| F\|_{\infty,\epsilon}\leq
\frac {1}{-\sigma}
\| G\|_{\infty,\epsilon}.
\label{8.3}
\end{equation}
Eq. (\ref{8.3}) shows that the irrelevant solution to the
renormalization group PDE is not only finite but also
decreases in the $L_{\infty,\epsilon}$-norm.

\subsection{Marginal case}

Let $\sigma =0$. In this case we assemble $F(p)$ using a
first order Taylor formula. Suppose then that we have
$L_{\infty,\epsilon}$-bounds on the first derivatives
\begin{equation}
\| G_{\mu}\|_{\infty,\epsilon}=
\sup_{p\in\R ^N} \left\{
\left\vert \frac{\partial}{\partial p^{\mu}} G(p) \right\vert
e^{-\epsilon |p|} \right\} <\infty.
\label{8.4}
\end{equation}
If $F(p)$ is marginal, then its first derivatives are
irrelevant with scaling dimension minus one. It follows that
\begin{equation}
\| F_{\mu}\|_{\infty,\epsilon} \leq
\| G_{\mu}\|_{\infty,\epsilon}.
\label{8.5}
\end{equation}
Therefrom it follows that
\begin{align}
|F(p)| e^{-\epsilon |p|} &\leq
|F(0)| e^{-\epsilon |p|}+
\sum_{\mu} |p_\mu |
\int_0^1 {\rm d}t
| F_\mu (tp)| e^{-\epsilon |p|}
\nonumber \\
&\leq
|F(0)|+
\sum_{\mu} |p_\mu |
\int_0^1 {\rm d}t
e^{-(1-t)\epsilon |p|}
\| F_\mu\|_{\infty,\epsilon}.
\label{8.6}
\end{align}
The result is an $L_{\infty,\epsilon}$-bound
\begin{equation}
\| F\|_{\infty,\epsilon}
\leq |F(0)|+
\frac {1}{\epsilon} \sum_{\mu}
\| F_\mu\|_{\infty,\epsilon}.
\label{8.7}
\end{equation}
This estimate is not uniform in $\epsilon$. It works for
$\epsilon$ arbitrary small, but the bound grows with an
inverse power of $\epsilon$. The large momentum growth
is a consequence of the split in derivatives and Taylor
remainder.

\subsection{Relevant case}

Let $\sigma >0$. This case requires a generalization of
the bound in the marginal case. The Taylor expansion is
pushed to order $\sigma +1$. Then the derivatives become
irrelevant. We assume $L_{\infty,\epsilon}$-estimates on
all derivatives
\begin{equation}
\| G_\alpha \|_{\infty,\epsilon} =
\sup_{p\in\R ^N} \left\{
\left\vert \frac{\partial ^{|\alpha |}}
{\partial p^{\alpha}} G(p)\right\vert
e^{-\epsilon |p|}\right\} <\infty
\label{8.8}
\end{equation}
of order $|\alpha |=\sigma +1$. Since they are irrelevant
with scaling dimension minus one it follows that the
corresponding derivatives of $F(p)$ obey
\begin{equation}
\| F_\alpha \|_{\infty,\epsilon}\leq
\| G_\alpha \|_{\infty,\epsilon},
\label{8.9}
\end{equation}
and are also $L_{\infty,\epsilon}$-bounded. From the
Taylor formula it then follows that
\begin{align}
|F(p)| e^{-\epsilon |p|} &\leq
\sum_{|\alpha |\leq\sigma}
\frac {|p^\alpha |}{\alpha !} e^{-\epsilon |p|}
| F_\alpha (0)|+
\sum_{|\alpha|=\sigma +1}
\frac {|p^\alpha |}{\alpha !}
\int_0^1{\rm d}t (1-t)^\sigma
|F_\alpha (tp)| e^{-\epsilon |p|}
\nonumber \\ &\leq
\sum_{|\alpha |\leq\sigma}
\frac {|p|^{|\alpha |}}{\alpha !} e^{-\epsilon |p|}
| F_\alpha (0)|+
\nonumber\\&\phantom{\leq}
\sum_{|\alpha|=\sigma +1}
\frac {|p|^{\sigma +1}}{\alpha !}
\int_0^1{\rm d}t (1-t)^\sigma
e^{-(1-t)\epsilon |p|}
\|F_\alpha \|_{\infty,\epsilon},
\label{8.10}
\end{align}
and thus
\begin{equation}
\| F \|_{\infty,\epsilon}\leq
\sum_{|\alpha |\leq \sigma} \frac {1}{\alpha !}
A_{\epsilon,|\alpha |} |F_{\alpha}(0)|+
\sum_{|\alpha |=\sigma+1} \frac {1}{\alpha !}
B_{\epsilon, \sigma+1} \| F_\alpha \|_{\infty,\epsilon},
\label{8.11}
\end{equation}
with constants
\begin{equation}
A_{\epsilon,|\alpha|}=\sup_{p\in\R ^N}
\left\{ |p|^{|\alpha |} e^{-\epsilon |p|}\right\},\quad
B_{\epsilon, \sigma +1}=\frac {\Gamma (\sigma +1)}
{\epsilon^{\sigma +1}}.
\label{8.12}
\end{equation}
Thus we again have an $L_{\infty,\epsilon}$-bound on
the function $F(p)$. This completes the large momentum bound
on $F(p)$. Exactly the same strategy applies to the derivatives
of $F(p)$ as well. The irrelevant derivatives inherit immediately
large momentum bounds. The relevant derivatives require Taylor
expansions. We omit to spell out explicitely the necessary
bounds on the derivatives of $G(p)$.

\subsection{Iteration and regularity}

The iterative scheme determines order by order $\beta^{(s)}$,
$\zeta^{(s-1)}$, $\mu^{(s)}$, and the irrelevant remainders
$\wt{\VV}^{(s)}_{irr,2n}(p_1,\ldots,p_{2n-1})$. It is finite
to all orders of perturbation theory because of the following
iteration of regularity. Suppose that we have shown the
following to all orders $s\leq r-1$: \\[2mm]
{\it I) $\beta^{(s)}$,
$\zeta^{(s-1)}$, and $\mu^{(s)}$ are finite numbers. II)
$\wt{\VV}^{(s)}_{irr,2n}(p_1,\ldots,p_{2n-1}$ is a
smooth function on $\R\times\cdots\times\R$ for all
$1\leq n\leq s+1$, symmetric in the momenta, and
$O(D)$-invariant. III)
$\|\wt{\VV}^{(s)}_{irr,2n,\alpha}\|_{\infty,\epsilon}$
is finite for all $\epsilon >0$, $1\leq n\leq s+1$, and
$|\alpha |\geq 0$. Here $\alpha$ is a multi-index which
labels momentum derivatives.} \\[2mm]
Then the same statements hold at order $s=r$. Since they
are trivially fulfilled to order one they iterate to all
orders of perturbation theory.

To prove the iteration of regularity we once more inspect
each step of the iterative scheme. First, the irrelevant
remainders $\wt{\KK}^{(r)}_{irr,2n}(p_1,\ldots,p_{2n-1})$
are smooth functions on $\R\times\cdots\times\R$,
symmetric under permutations and $O(D)$-invariant. They
and all their momentum derivatives satisfy
$L_{\infty,\epsilon}$-bounds. They are composed of two
contributions. The first immediately inherits a bound from
the induction hypothesis. The second is a sum of
renormalization group brackets of lower orders.
Therefore, they consist of multiple convolutions with
propagators. The integrals converge, are smooth functions
of the external momenta, and satisfy
$L_{\infty,\epsilon}$-bounds. Second, we have linear
equations for the coefficients $\beta^{(r)}$,
$\zeta^{(r-1)}$, and $\mu^{(r)}$ with finite coefficients.
Third, the integration of the inhomogeneous renormalization
group PDEs, yields solutions with the desired properties.

\section{Conclusions}

The aim of perturbative renormalization theory is to derive
power series expansions for Green's functions which are free
of divergencies. The BPHZ theorem states that this can be
accomplished by writing the Green's functions in terms of
renormalized parameters. An elegant proof of the BPHZ theorem
was given by Callan \cite{C76}. A polished version of
which is due to Lesniewski \cite{L83}. Their method is
similar to ours in that it is based on renormalization group
equations for the renormalized Green's functions, the
Callan-Symanzik equations. The method proposed here is different
in that it does {\it not} resort to any kind of graphical
analysis, not to analysis of sub-graphs, and not to skeleton
expansions.

A new generation of proofs of the BPHZ theorem was
initiated with the work of Polchinski \cite{P84}. His proof
has been simplified further by Keller, Kopper, and
Salmhofer \cite{KKS90}. Our method is similar in that
it uses an exact renormalization differential equation.
The details are however quite different. The main difference is
that Polchinski begins with a cutoff theory. He then shows how
the cutoff can be removed in a way such that the effective
interaction remains finite. Our method directly addresses the
limit theory without cutoffs, expressed in terms of a
renormalization group transformation with cutoffs.
Roughly speaking, we are here simultaneously changing Polchinski's
renormalization conditions and integrating an amount of
fluctuations. Unlike Polchinski and followers we use a
renormalization group differential equation with dilatation
term. A way to think of (\ref{1.5}) is as an equation for a
renormalization group fixed point of a system which has been
enhanced by one degree of freedom, the running coupling. This
fixed point problem can only be formulated with rescaling and with
dilatation term.

Another renormalization group approach to renormalized
perturbation theory comes from Gallavotti \cite{G85,GN85}
and collaborators. Pedagocial accounts of tree expansions
can be found in \cite{BG95,FHRW88}.
There the result of renormalization is
expressed in terms of a renormalized tree expansion. The
program of \cite{Wi96} with an iterated transformation
with fixed $L$ is related to the tree expansion. Both are
built upon a cumulant expansion for the effective interaction.
The renormalization procedure is however quite different.
Like Polchinski, Gallavotti starts from a cutoff theory.
It is organized in terms of trees, which describe the
sub-structure of divergencies in Feynman diagrams. The
divergencies are transformed into a flow of the non-irrelevant
couplings. This part is similar to ours. The basic difference
with Gallavotti is that we do not organize our
expansion in terms of trees. A hybrid approach between
Polchinski and Gallavotti is due to Hurd \cite{H89}.

An important question is whether this constuction of
renormalized trajectories extends beyond perturbation
theory.\footnote{It certainly works in the cases where
perturbation theory converges.} Another important question
is whether it extends to renormalized trajectories at
non-trivial fixed points. We hope to return with answers
to these questions in the future.


\begin{thebibliography}{XXXXX}
%
\bibitem[BG95]{BG95} G.\,Benfatto, G.\,Gallavotti,
Renormalization group, Physics Notes No. 1, Princeton
University Press 1995
%
\bibitem[C76]{C76} C.\,G.\,Callan, Introduction to
renormalization theory, Les Houches Lecture Notes
1975, 41-77, R.\,Balian and J.\,Zinn-Justin eds.
%
\bibitem[FHRW88]{FHRW88} J.\,S.\,Feldman, T.\,R.\,Hurd,
L.\,Rosen, J.\,D.\,Wright, QED: A proof of renormalizability,
Lecture Notes in Physics 312, Springer Verlag 1988
%
\bibitem[G85]{G85} G.\,Gallavotti, Renormalization theory
and ultraviolet stability for scalar fields via renormalization
group methods, Rev. Mod. Phys. Vol. 57 No. 2 (1985) 471-562
%
\bibitem[GN85]{GN85} G.\,Gallavotti and F.\,Nicolo,
Renormalization in four dimensional scalar fields I,
Commun. Math. Phys. 100 (1985) 545-590; Renormalization
Renormalization in four dimensional scalar fields II,
Commun. Math. Phys. 101 (1985) 247-282
%
\bibitem[GJ87]{GJ87} J.\,Glimm and A.\,Jaffe, Quantum Physics,
Springer Verlag 1987
%
\bibitem[H89]{H89} T.\,Hurd, A renormalization group proof
of perturbative renormalizability, Commun. Math. Phys. 124
(1989) 153-168
%
\bibitem[KKS90]{KKS90} G.\,Keller, C.\,Kopper, M.\,Salmhofer,
Perturbative renormalization and effective Lagrangeans,
MPI-PAE/PTH 65/90
%
\bibitem[L83]{L83} A.\,Lesniewski, On Callan's proof of the
BPHZ theorem, Helv. Phys. Acta, Vol. 56 (1983) 1158-1167
%
\bibitem[P84]{P84} J.\,Polchinski, Renormalization and
effective Lagrangeans, Nucl. Phys. B231 (1984) 269-295
%
\bibitem[W71]{W71} K.\,Wilson, Renormalization group and
critical phenomena I and II, Phys. Rev. B4 (1971) 3174-3205
%
%
%
\bibitem[Wi96]{Wi96} C.\,Wieczerkowski, The renormalized
$\phi^4_4$-trajectory by perturbation theory in the running
coupling I: the discrete renormalization group, hep-th/9601142
%
\bibitem[WK74]{WK74} K.\,Wilson and J.\,Kogut,
The renormalization group and the $\epsilon$-expansion,
Phys. Rep. C12 No. 2 (1974) 75-200
%
\end{thebibliography}
\end{document}